\documentclass[aps,preprint,amsmath,amssymb]{revtex4}
\usepackage{graphicx}
\def\be{\begin{eqnarray}}
\def\en{\end{eqnarray}}

\begin{document}
\title{\Large \bf Dilepton decays and oscillation of $B_{s}$
 in split supersymmetry with R-parity violation}

\date{\today}
\author{\large \bf  Chuan-Hung~Chen$^{a,b}$\footnote{Email:
phychen@mail.ncku.edu.tw} and Chao-Qiang~Geng$^{c}$ \footnote{Email:
geng@phys.nthu.edu.tw}   }
 \affiliation{$^{a}$Department of Physics,
National Cheng-Kung University, Tainan, 701
Taiwan\\
$^{b}$National Center for Theoretical Sciences, National Cheng-Kung
University, Tainan, 701
Taiwan\\
$^{c}$Department of Physics, National Tsing-Hua University, Hsin-Chu
, 300 Taiwan }

\begin{abstract}
We study $B$ physics phenomenology in the scenario of split
supersymmetry without R-parity. By assuming the constraints of
bilinear (trilinear) R-parity violating couplings, which are
introduced to solve the problem of  the  atmospheric (solar)
neutrino mass, we show that the decay branching ratios of
$B_{s}\rightarrow \ell^{+} \ell^{-}$ and the mixing of
$B_s-\bar{B}_s$ can be large. Explicitly, we find that
$B(B_s\to\mu^+\mu^-)=O(10^{-7})$ and $\Delta m_{B_s}=O(10^{-9})\
GeV$, which should be observed at future hadron colliders.

\end{abstract}
\maketitle

It is believed that the standard model (SM) is not complete due to
most phenomena being based on $19$ input parameters \cite{DGH}. It
is expected that new physics should exist at some
high energy scale $\Lambda$ to smear the divergent mass of Higgs,
induced from one-loop level. Otherwise, the principle of
naturalness \cite{thooft} is breakdown while $\Lambda $ goes to
the scale which is much higher than that of electroweak.
%, where the SM suffers from the fine tuning problem
%in Higgs sector.
It is found
that extending the SM to supersymmetry (SUSY)
at the scale $\Lambda$ of
$O(\rm TeV)$ can solve not only the hierarchy problem, but
also the problem of unified gauge coupling \cite{unify1,unify2}.
Furthermore, the predicted lightest neutralino in supersymmetric
 models could also provide the candidate of dark matter
\cite{unify1,dark}.

%Although the introduced supersymmetry succeeds in solving the main
%theoretical problem,
Apart from the above successes, models with SUSY
still suffer some
difficulties from phenomenological reasons, such as the problems on
small CP violating phases, large flavor mixings and proton decays,
as well as
%. However, the most serious one is the
they predict too large
cosmological constant. Inevitably, fine tuning always appears in
the low energy physics. Recently, in order to explain the
cosmological constant problem and preserve the beauty of
the ordinary low-energy SUSY models, the scenario of split SUSY
is suggested \cite{AD,ADGR}, in which  the SUSY breaking scale is
much higher than the electroweak scale.
In this split SUSY scenario,
 except the SM Higgs which
could be as light as the current experimental limit, the scalar
particles are all ultra-heavy. On the other hand, by the
protection of approximate chiral symmetries, the masses of
fermions, such as gauginos and higgsinos, could be at the
electroweak scale \cite{AD,GR}.

Based on the aspect of split SUSY, various interesting topics
on particle physics phenomenology have been studied, including,
for instance, physics at colliders \cite{Zhu},  Higgs
 \cite{Mahbubani}, phenomena of stable gluino \cite{Hewett},
 sparticles in cosmic rays \cite{AGN}, dark matter
 \cite{GKM,AJM},  grand unified theories (GUTs) \cite{sakar},  neutrino
physics \cite{CP}, and so on.

In this paper, we examine the implication of the split SUSY scenario
on B physics at hadron colliders, such as BTeV and LHC. In
particular, we explore the possibility of having large effects in
the dilepton decays and oscillation in the $B_s$ system with split
SUSY. In the conventional SUSY models with R-parity invariance, it
is known that the gaugino penguin and box diagrams have significant
contributions to $B$ processes, such as $B_{d(s)}-\bar{B}_{d(s)}$
mixings, $B\to X_{s} \gamma$ \cite{GGMS}, $B\to K^* \ell^{+}
\ell^{-}$ \cite{CGPRD66}, and the time-dependent CP asymmetry of
$B\to \phi K_{S}$ \cite{CG} etc. However, since the diagrams
involved are associated with squarks in the internal loop, the
results in the split SUSY  will be highly suppressed by the masses
of squarks, denoted by $m_{S}$. Hence, one suspects that the
scenario of split SUSY with R-parity could not induce interesting
phenomena from low energy physics.

In this study, we consider split SUSY in the framework of R-parity
violation. It has been pointed out recently in Ref. \cite{GKM} that
the lightest neutralino in the R-parity violating model could still
remain the candidate of dark matter. Although R-parity violation
leads to the decay of neutralino, by the suppression of the
high-scale SUSY breaking, the neutralino lifetime could exceed the
age of our universe. Moreover, by the combination of bilinear and
trilinear couplings, it has been shown in Ref. \cite{CP} that the
observed mass scales of atmospheric and solar neutrinos can be
accommodated in the split SUSY scenario without R-parity.
%%%%%%%%%%%%%%%%%%%%%%%%%%%%%%%%%%%%%%%%%%%%%%%%%%%%%%%%%
%We note that although one doesn't need to employee SUSY to solve the
%problem of neutrino masses, in order to provide the candidate of
%dark matter, the solution to cosmological constant and new effects
%on low energy physics etc consistently, we still study the physics
%in the framework of SUSY.
%%%%%%%%%%%%%%%%%%%%%%%%%%%%%%%%%%%%%%%%%%%%%%%%%%%%%%%%
%%%%%%%%%%%%%%%%%%%%%%%%%%%%%%%%
 In our analyses, we will
assume that the neutrino mixing arises from
the neutralino-neutrino mixing in our split SUSY scenario.
However, it should be clear that one does
not need SUSY to explain neutrino masses or mixings,
which can be understood quite well in
the context of the standard model, with appropriate assumptions on
neutrino mass matrix.
%%%%%%%%%%%%%%%%%%%%%%%%%%%%%%%%
Based on the results in Refs. \cite{GKM,CP}, we find that
% in the split SUSY scenario,
if the induced mixing of sneutrino and Higgs is not constrained by the
neutrino masses, the decay BR of $B_{s}\to \mu^{+} \mu^{-}$ can
reach the current experimental upper limit, whereas the effect of
the $B_s-\bar{B}_s$ oscillation can be two orders larger than that
in the SM.

We start by introducing the interactions of R-parity violating
terms. In terms of the notations in Ref. \cite{CP} , the bilinear
and trilinear terms for the lepton number violation in the
superpotential can be written as \cite {CP,CJP,KO}
 \be
 W=\mu H_1 H_2+\epsilon_i \mu L_i H_2+\lambda^{\prime}_{ijk} L_i Q_j
 D^{c}_{k} + \lambda_{ijk} L_{i} L_{j} E^{c}_{k}, \label{eq:w}
  \en
and the relevant scalar potential is given by
   \be
   V=BH_1H_2+B_{i}L_{i}H_{2}+m^{2}_{L_i H_1}L_i H^{\dagger}_{1}
   +h.c.\, . \label{eq:v}
   \en
Note that, for simplicity, we have neglected the baryon number
violating effects and used the same notations for superfields and
ordinary fields. In split SUSY, the soft parameters $B$, $B_{i}$ and
$m^{2}_{L_i H_1}$ could be the order of $m^{2}_{S}$. It is known
that $m_S$ is in the range of $10^{9}-10^{13}$ GeV
\cite{AD,ADGR,GR}. From Eqs.~(\ref{eq:w}) and (\ref{eq:v}),
the bilinear R-parity violating terms can make the vacuum
expectation values (VEVs) of sneutrino fields be nonzero \cite{vev}.
In terms of a set of tadpole equations \cite{tadpole} which are the
conditions for obtaining the stable potential, these VEVs
 are given by $\tilde{v}_{i}=\left|\langle H^{0}_{2}
\rangle B_{i} +\langle H^{0}_{1}\rangle m^{2}_{L_i
H_1}\right|/m^{2}_{L_i}$, where we have neglected
the small contributions of D terms for simplicity. Clearly, if $B_{i}$ and
$m^{2}_{L_i H_1}$ are the same order of magnitude or smaller than
the masses of sleptons, the VEVs are still satisfied with
the conditions of minimum potential. By the couplings of
slepton-lepton-gaugino, neutrinos and charged leptons will mix with
neutralinos and charginos. Consequently, they induce neutrino
masses at tree and loop levels, respectively.
% while the latter induces by loop effects.
%The relevant constraints by
%neutrino masses can be found in Ref.~\cite{DMR}.
It has been shown \cite{CP} that to explain the atmospheric neutrino
mass scale $\sqrt{\Delta m^2_{atom}} \sim 0.05$ eV, the involving
parameters, associated with bilinear couplings and defined by
$\xi_{i}=\tilde{v}_i/\langle H^0_1 \rangle -\epsilon_i=m^2_{L_i
H_1}/m^{2}_{L_i}+B_i/m^{2}_{L_i} \tan\beta -\epsilon_i$ with
$\tan\beta=\langle H^{0}_{2}\rangle / \langle H^0_1 \rangle$, are
limited to $10^{-6}/\cos\beta$ at tree level. That is, a strong
cancelation occurs among R-parity violating parameters in $\xi_{i}$.
Since  the fine-tuning problem has been abandoned in split SUSY,
each term in $\xi$ could be order of unity. In order to obtain the
mass scale of solar neutrino, $\sqrt{\Delta m^2_{sol}} \sim 9$ meV,
one has to go to one-loop level, induced by the same bilinear
couplings. However, the results are suppressed by $\eta_{i}
m^{2}_{Z}/m^{2}_{L_i}$ with $\eta_i=\xi_i-B_i/B$ \cite{CP,CJP}. To
solve the solar neutrino mass problem, it is concluded \cite{CP}
that the trilinear R-parity violating couplings
$\lambda^{\prime}_{i23}$ and $\lambda^{\prime}_{i32}$ need to be of
order one.

\begin{figure}[hpbt]
%\includegraphics*[width=1.6
%in]{one-loop}
\includegraphics*[width=2.5in]{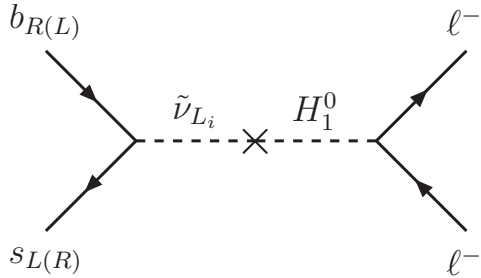} \caption{Tree contribution
to $B_s\to\ell^+\ell^-$ with the cross representing the mixings
between sleptons and Higgs.}
 \label{tree1}
\end{figure}

In the following discussions, we will take that
$\lambda^{\prime}_{i23,i32}$ as well as the ratios of the bilinear
couplings and $m_S^2$, i.e., $m^{2}_{L_i H_1}/m^2_{S}$ and
$B_i/m^2_{S}$, are order of unity. It is interesting to
investigate if there are some observable physics phenomena beside
those discussed in Ref. \cite{CP}. Since
$\lambda^{\prime}_{i23,i32}\sim 1$, it is natural for us to think
of physics involving the flavor changing natural current (FCNC) of
 $b\to s $ transition. Indeed, we find that
 $B_{s} \to \ell^{+} \ell^{-}$ can occur at tree level as shown
in Fig. \ref{tree1}, which may not be suppressed.
%%%%%%%%%%%%%%%%%%%%%%%%%%%%%%%%%%%%%%%%%%%%%%%%%%%%%%%%%%%%%
In the SM, it is known that $B_s\to \ell^{+} \ell^{-}$ decays arise
from the electroweak penguin and box diagrams. The decay branching
ratio (BR) of  $B_s\to \mu^{+} \mu^{-}$ is found to be $(3.8 \pm
1.0)\times 10^{-9}$ \cite{BurasPLB} which is much less than the
current experimental upper limit of $5.0\times 10^{-7}$ \cite{D0}.
From the relationship $B(B_{s}\to \tau^{+} \tau^{-})/B(B_s\to \mu^+
\mu^-) =\bar{m}^2_{\tau}/\bar{m}^2_{\mu}$ with
$\bar{m}_{\ell}=m_{\ell}(1-4m_{\ell}^2/m_{B_s}^2)^{1/2}$
\cite{Bsll}, the corresponding tau mode can be also studied. It was
demonstrated that $B(B_s\to \mu^{+}\mu^{-})=O(10^{-7})$  could be
achieved in ordinary SUSY models \cite{KKL}. However, it is easy to
check that these contributions are suppressed in the split SUSY
scenario. Since in the split SUSY approach, except the SM-like Higgs
denoted by $h^0$ is light, all scalars are extremely heavy.
Therefore, we may simplify the calculations by using $-h^0
\sin\alpha\, (h^0 \cos \alpha)$ instead of the Higgs $H^{0}_{1}
(H^0_{2})$, where the angle $\alpha$ describes the mixing of two
neutral Higgses \cite{Higgs}. In terms of the interactions in Eqs.
(\ref{eq:w}) and (\ref{eq:v}), the decay amplitude for $B_{s}\to
\ell^{+} \ell^{-}$ is given by
    \be
       A=\langle \ell^{+} \ell^{-} | H_{eff} |\bar{B}_s\rangle =
       \frac{-i}{2m^{2}_{h}}\left(\frac{gm_{\ell}}{2m_{W}} \frac{\sin\alpha }{\cos\beta}\right)
       \frac{f_{B_s} m^{2}_{B_s} }{(m_b+m_s)}\frac{m^{2}_{L_{i}
       H_1} (\lambda^{\prime*}_{i23}-
       %m^{2*}_{L_{i}H_1}
        \lambda^{\prime}_{i32})}{m^{2}_{\tilde{\nu}_i}}\, \bar{\ell}\, \ell
       %=C_{\ell}
       %\, \bar{\ell}\, \ell,
       \label{eq:bll}
    \en
where $m_h$, $m_W$, $m_{b}$, $m_s$, $m_{\ell}$, $m_{B_s}$ stand
for the masses of Higgs, W-boson, b-quark, s-quark, lepton and
$B_{s}$, respectively. As mentioned early, except the SM-like Higgs,
the masses of sfermions, scalar and pseudoscalar bosons are much
higher than the electroweak scale. Therefore,
%we neglect to include
the contributions from other scalar particles will be neglected. In
Eq.~(\ref{eq:bll}), the second factor and $m^2_{L_iH_1}$ are from
the coupling of the SM Higgs to the lepton and the mixings between
sleptons and Higgs, respectively. In the equation, we have also used
the identity $\langle0 | \bar{s} \gamma_5
b|\bar{B}_s\rangle\approx -i\, f_{B_s} m^2_{B_s}/(m_b+m_s)$. Since
the trilinear couplings in sleptons and quarks involve two
possible chiralities, there is a cancelation in
%the last factor of
Eq.~(\ref{eq:bll}). Note that if
%the couplings are real and
$\lambda^{\prime*}_{i23}=\lambda^{\prime}_{i32}$, our mechanism
vanishes automatically.
%%%By using $\sum (\bar{\ell}{\ell})(\bar{\ell}
%%\ell)^{\dagger}=2(m^{2}_{B_s}-4m^2_{\ell})$.
By squaring the decay amplitude and including the phase space
factor, the decay rate is derived to be
       \be
         \Gamma&=&
         \frac{m_{B_s}}{16\pi }
         \frac{G_{F} m^{2}_{\ell}}{\sqrt{2}}  \left( \frac{f_{B_s}m_{B_s}
         }{m^{2}_{h}} \frac{\sin\alpha}{\cos\beta}
        |{\cal N}_i| \right)^2
          \left[ 1- \left( \frac{2m_{\ell}}{m_{B_s}} \right)^2
         \right]^{3\over 2}
       \label{Rate}
       \en
with ${\cal
N}_i=m^{2}_{L_{i}H_1}(\lambda^{\prime*}_{i23}-
\lambda^{\prime}_{i32})/ m^{2}_{\tilde{\nu}_i}$. We note that the
decay rate is proportional to $m^{2}_{\ell}$, which is the same
as that in the SM.

\begin{figure}[hpbt]
%\includegraphics*[width=1.6
%in]{one-loop}
\includegraphics*[width=3.in]{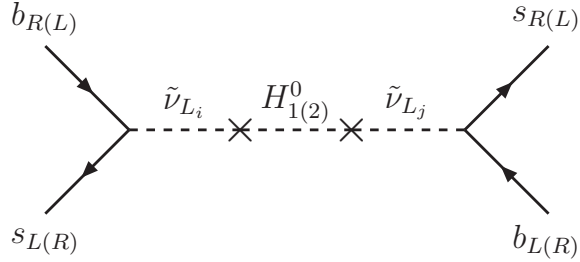} \caption{Tree contribution
to the mixing of $B_s-\bar{B}_s$ with the crosses representing the mixings
between sleptons and Higgs.}
 \label{tree2}
\end{figure}

Besides $B_s\to \ell^+ \ell^-$ decays, we find that the same
mechanism can also generate other FCNC processes, such as the
$B_{s}-\bar{B}_{s}$ mixing, induced by the W-exchange box
diagrams in the SM. We note that its SM value is
$(1.19\pm0.24)\times 10^{-11}$ GeV \cite{Barger}, while the
current experimental limit is  larger than $9.48\times 10^{-12}$
GeV \cite{PDG}. In SUSY models with R-parity, the main effects are
also from the box diagrams but with gluinos and charginos instead
of W-boson in the loops \cite{GGMS}. Unfortunately, the resultants
are associated with $1/m^{2}_{S}$, which are obviously highly
suppressed in split SUSY. However, if we insert one more mixing of
sneutrinos and Higgses in the Higgs propagator of Fig.
\ref{tree1}, the $B_s-\bar{B}_s$ oscillation could be induced at
tree-level too, as shown in Fig. \ref{tree2}. Consequently, the
effective Hamiltonian is obtained as
   \be
     {\cal H}_{eff}&=&
     %\frac{\lambda^{\prime *}_{i23}\lambda^{\prime}_{j32}}{m^{2}_{h}}
     %  (\bar{s}\, P_R \,
    % b) (\bar{s}\, P_L \, b)+h.c.\, ,
 \frac{\lambda^{\prime *}_{i23}\lambda^{\prime}_{j32}}{m^{2}_{h}}
     {\cal C}_{ij}  (\bar{s}\, P_R \,
     b) (\bar{s}\, P_L \, b)+h.c.\, ,
\label{eq:heff}
   \en
        where
    \be
       {\cal C}_{ij} &=&\frac{
       %\lambda^{\prime *}_{i23}\lambda^{\prime}_{j32}
       1}{m^{2}_{\tilde{\nu}_i}m^{2}_{\tilde{\nu}_j}}
      \left[ B_i B_j^*
     \cos^2\alpha
     +  m^{2}_{L_i H_1}m^{2}_{L_j H_1}
     \sin^2\alpha\right]\,.
     % (\bar{s}\, P_R \,
     %b) (\bar{s}\, P_L \, b)+h.c.\, ,\nonumber\\
     \label{eq:heff1}
     \en
     It is interesting to note that if $B_{i}=m^{2}_{L_i H_1}$, ${\cal
     C}_{ij}$ will be independent of the angle $\alpha$.
   %%%%%%%%%%%%%%%%%%%%%%%%%%%%%%%%%%%%%%%%%%%%%%%%%%%%%%%%%%%%%
 From Eq. (\ref{eq:heff}), we see that the induced oscillation is
associated with the multiple of $\lambda^{\prime*}_{i23}
\lambda^{\prime}_{j32}$. By considering the CP conserving case, the
effective couplings are
 similar to those for the
solar neutrino masses, presented by \cite{CP}
  \be
   M^{\nu}_{ij}\sim \frac{3}{8\pi^2}\lambda^{\prime}_{i23}
   \lambda^{\prime}_{j32}\frac{m_b m_s}{m_S}. \label{eq:neutrino}
  \en
To estimate the hadronic matrix element, we employ the results of
vacuum insertion method, given by \cite{GGMS}
      \be
         \langle \bar{B}_{s} |\bar{s}\, P_R\, b\, \bar{s}\, P_L\, b|
         B_{s}\rangle\approx \left[\frac{1}{24}+ \frac{1}{4} \left ( \frac{m_{B_s}}{m_b+m_s}\right)^2\right]
          m_{B_s} f^2_{B_s}.
      \en
As a result, the mass difference for $B_s$ and $\bar{B}_s$  is
described by
      \be
      \Delta m_{B_s}=2  |M_{12}|=\frac{4}{m^{2}_{h}}
       \left|Re(\lambda^{\prime *}_{i23}\lambda^{\prime}_{j32}{\cal C}_{ij})\right|
       \left[\frac{1}{24}+ \frac{1}{4}
        \left ( \frac{m_{B_s}}{m_b+m_s}\right)^2\right] m_{B_s}
        f^2_{B_s}.
    \label{Dmbs}  \en
Hence, it will be interesting to see if large contributions on BRs of
$B_{s}\to \ell^{+}\ell^{-}$ and the $B_s$ oscillation can be
obtained  in our split SUSY scenario.

\begin{figure}[hpbt]
\includegraphics[width=2.5in]{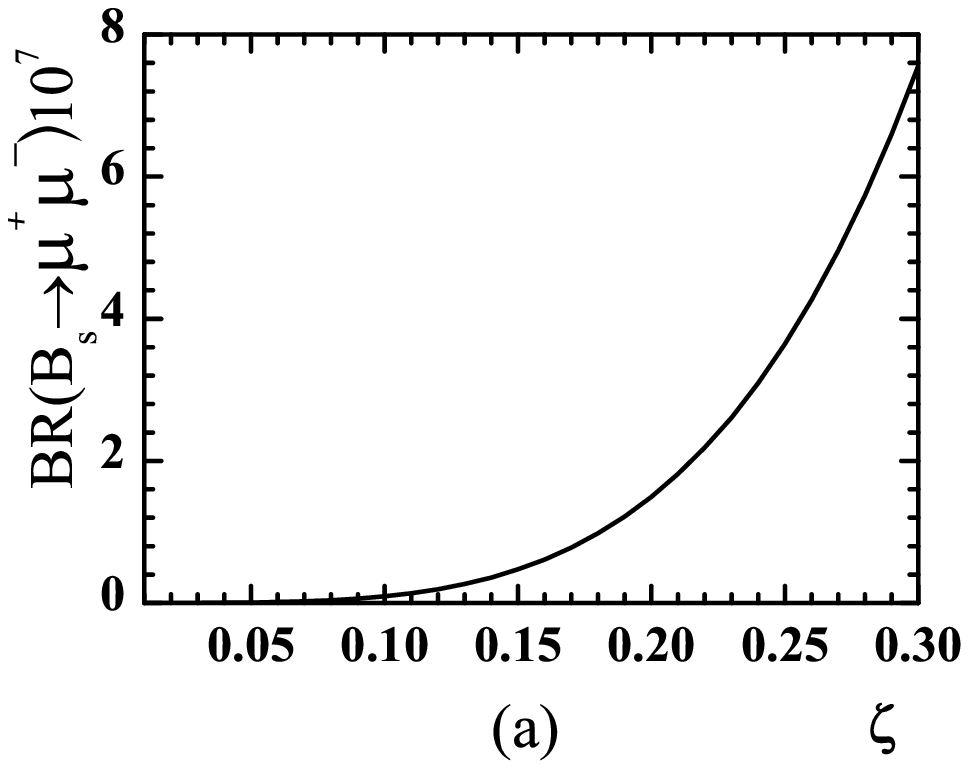}  \includegraphics[width=2.5in]{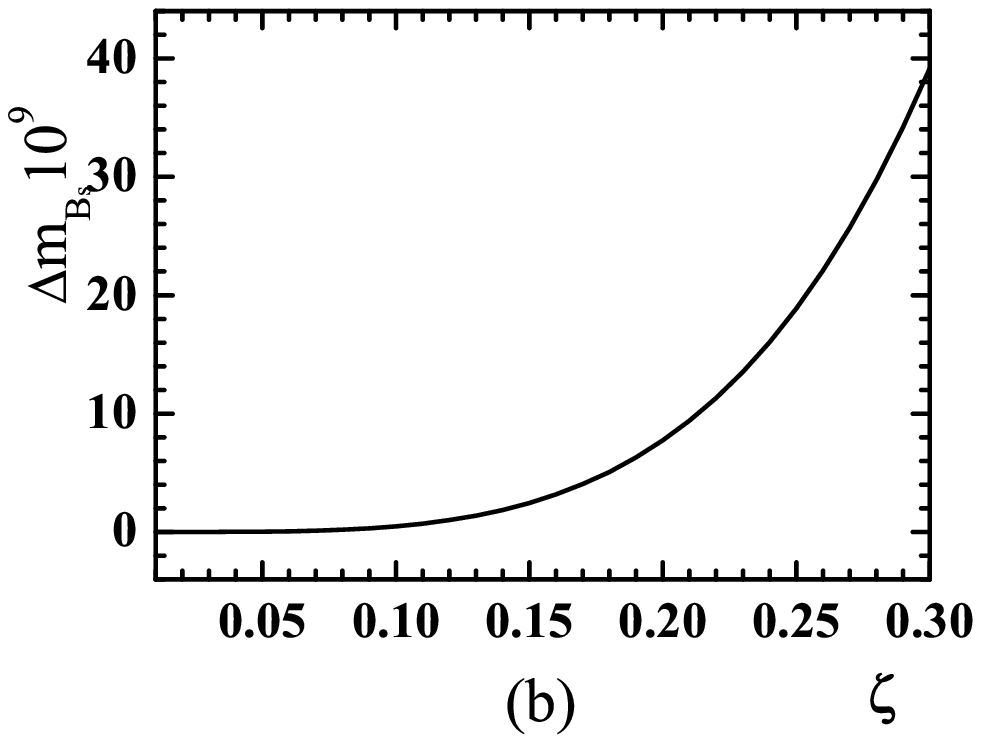}
\caption{ (a) BR($B_s\to \mu^{+} \mu^{-})$ and (b) $\Delta m_{B_s}$
as functions of $\zeta$ with $m_{h}=150$ GeV.}
%
%The solid, dashed, and dash-dotted
%lines in (a) denote $\tan\beta=1$, $5$ and $10$, respectively.}
 \label{fig:br-mixing1}
\end{figure}

\begin{figure}[hpbt]
%\includegraphics*[width=1.6
%in]{one-loop}
\includegraphics[width=2.5in]{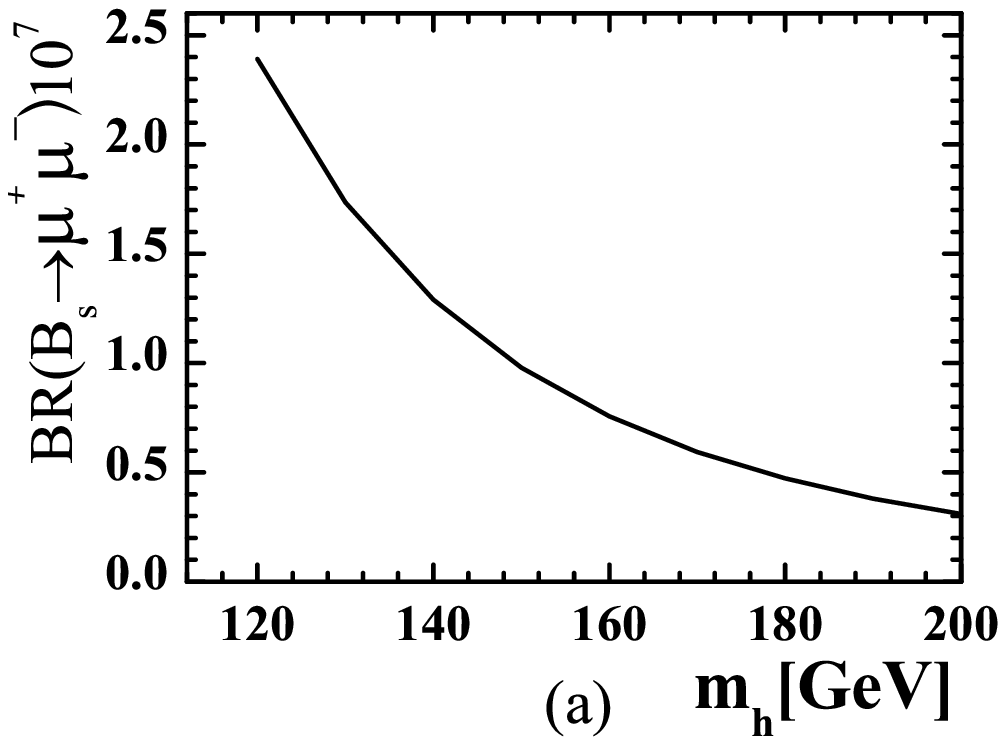}  \includegraphics[width=2.38in]{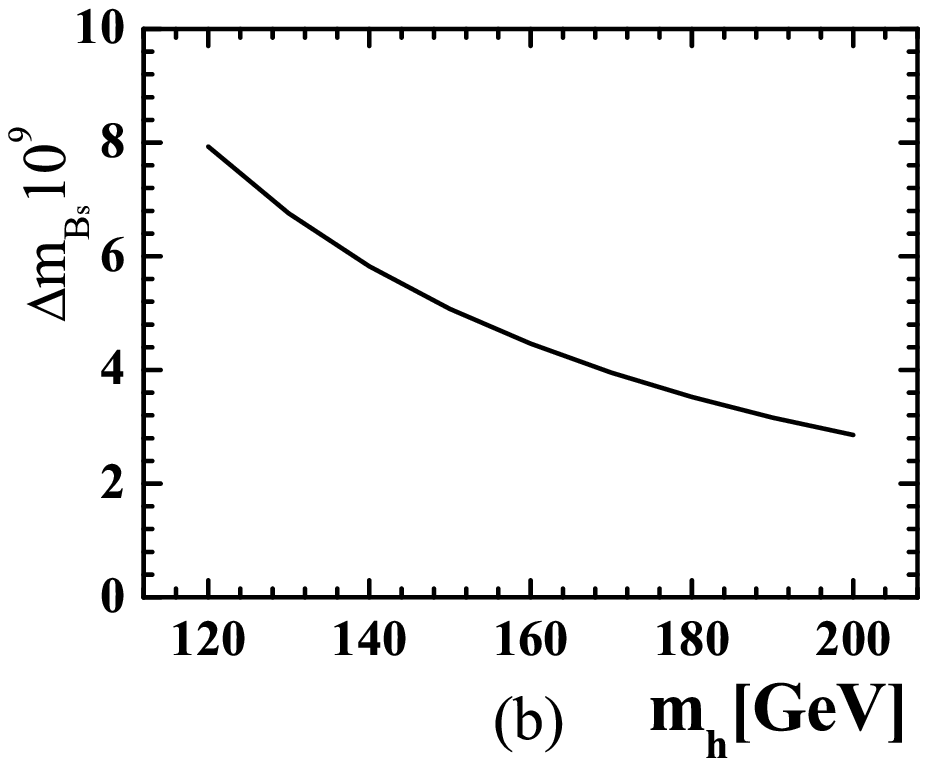}
\caption{ (a) BR($B_s\to \mu^{+} \mu^{-})$ and (b) $\Delta m_{B_s}$
as functions of $m_h$ with $\zeta=0.18$.}
% The Legend is the same as
%Fig.~\ref{fig:br-mixing1}.}
 \label{fig:br-mixing2}
\end{figure}

To estimate the numerical values, we take $f_{B_s}=0.23$ GeV
\cite{BBP}, $m_{B_s}=5.37$ GeV , $m_b=4.5$ GeV,  $m_s=0.13$ GeV, and
$\tau_{B_s}=1.46\times 10^{-12}\, s$ \cite{PDG}. In order to
preserve the solar neutrino mass to be $\sim 9$ meV, we set
$\lambda^{\prime}_{i23}=0.9$ and $\lambda^{\prime}_{j32}=-0.3$.
As one of CP-even Higgs bosons is very heavy,
%the angle $\alpha$ could be chosen as
 $\alpha\simeq\pi/2+\beta$ and $\sin\alpha/\cos\beta=1$.
 Therefore, in the split SUSY scenario,
%  Since the BR of $B_s\to\mu^{+} \mu^{-}$ is proportional to $\sin\alpha/%\cos\beta$, by the chosen angle $\alpha$,
 we see that the BR of $B_s\to\mu^{+} \mu^{-}$ is independent of the
angles $\alpha$ and $\beta$ due to due to Eq. (\ref{Rate}).
 For
simplicity, we set $\zeta=m_{L_i H_1}/m_{\tilde{\nu}_i}
  =\sqrt{|B_i|}/m_{\tilde{\nu}_i}$ so that  in our numerical estimations $C_{ij}=\zeta^{4}$.
To illustrate the specific values for $BR(B_{s}\to \mu^{+} \mu^{-})$
and $\Delta m_{B_s}$, by using Eqs. (\ref{Rate}) and (\ref{Dmbs})
and choosing $m_{h}=150$ GeV and $\zeta=0.18$, we get
  \be
   BR(B_s\to \mu^{+} \mu^{-})&=&1.0\times 10^{-7}\,, \nonumber \\
    \Delta  m_{B_s}&=& 4.8\times 10^{-9}\ {\rm GeV}\,.
\label{results}
  \en
We note that in Eq. (\ref{results}) the decay BR of
$B_s\to\mu^+\mu^-$ is close to the current experimental limit, while
$\Delta m_{B_s}$ is two orders of magnitude larger than the SM
prediction. It is clear that our results on $BR(B_{s} \to
\mu^{+}\mu^{-})$ and $\Delta m_{B_s}$ can be observed at hadron
colliders, such as BTeV and LHC which produce more than $10^{8}\
B_s\bar{B}_s$. In Figs. \ref{fig:br-mixing1} and
\ref{fig:br-mixing2}, we present $BR(B_s\to \mu^{+} \mu^{-})$ and
$\Delta  m_{B_s}$ as functions of $\zeta$ with $m_{h}=150$ GeV and
$m_{h}$ with $\zeta=0.18$, respectively.
   Since we have taken $B_{i}=m^{2}_{L_i H_1}$, the
  values of $\Delta m_{B_s}$ will not depend on angle $\alpha$.
%  From
%  Fig.~\ref{fig:br-mixing1}, we see clearly that $BR(B_s\to
%  \mu^{+}\mu^{-})$ is  very sensitive to not only $\tan\beta$ but
%  also $\zeta$, whereas $B_{s}$ is only sensitive to $\zeta$.
%%%%%%%%%%%%%%%%%%%%%%%%%%%%%%%%%%%%%%%%%%%%%%%%%%%%%%%%%%%%%
%%%%%%%%%%%%%%%%%%%%%%%%%%%%%%%
%%%%%%%%%%%%%%%%%%%%%%%%%%%%%%%%%%%%%%%%%%%%%%%%%%%%%%%%%%%%%

Finally, we remark that our mechanism could also
be used to the $B_{d}$ processes.  By using
$\lambda^{\prime}_{i13,j31}$ instead of
$\lambda^{\prime}_{i23,j32}$, similar phenomena will occur in
$B_{d} \to \ell^{+} \ell^{-}$ decays and the oscillation of
$B_{d}$. However, since $m_{d}<<m_{s}$, even with $\lambda^{\prime}_{i13,j31} $
being order of unity, there are no interesting contributions
to the solar neutrino masses. Moreover,
the $B_{d}-\bar{B}_{d}$ mixing could be used as the constraint on the corresponding trilinear
couplings.
In addition, it is worth mentioning that the tree contribution of
$b\to s \ell^{+} \ell^{-}$ in Fig. \ref{tree1} could lead large
effects on physics in $B\to K^{(*)}\ell^+\ell^-$ \cite{CG1} and
$\Lambda_b\to\Lambda\ell^+\ell^-$ \cite{CG2}. Similar conclusions
could be applied to $\tau^\pm\to\mu^\pm\mu^+\mu^-$ as well. The study will be presented elsewhere \cite{CG3}.

In summary, we have studied the implications of split SUSY on the
FCNC processes due to the $b\to s$ transition. It has been shown
that when the solar neutrino mass problem is solved in split SUSY
scenario, we find that the mixing effects of sneutrino and Higgs
could have large contributions to the BRs of $B_{s}\to \ell^{+}
\ell^{-}$ and the $B_s-\bar{B}_s$ mixing.
%Moreover, the similar
%mechanism could also have interesting effects on FCNC of $b\to d$
%transition. Once the trilinear couplings
%$\lambda^{\prime}_{i13,i31}$ are constrained by $\Delta m_{B_d}$,
%one could extent the contributions to dilepton decays of $B_{d}$.
\\

{\bf Acknowledgments}\\

We thank Kingman Cheung for helpful discussions. This work is
supported in part by the National Science Council of R.O.C. under
Grant \#s:  NSC-93-2112-M-006-010 and NSC-93-2112-M-007-014.

\end{document}